\documentclass[prl,twocolumn,superscriptaddress,floats]{revtex4}
\usepackage{graphicx}
\usepackage{amsmath}
\usepackage{booktabs}
\usepackage{dcolumn}
\usepackage{textcomp}

\usepackage{rotating}

\newlength{\La} 

\newlength{\Lb} 

\newlength{\Lc} 

\newcolumntype{d}{D{.}{.}{-1}}

\newcommand{\srruo}{Sr$_2$RuO$_4$}

\newcommand{\srdzs}{Sr$_3$Ru$_2$O$_7$}

\newcommand{\caruo}{Ca$_2$RuO$_4$}

\newcommand{\csruo}{Ca$_{2-x}$Sr$_{x}$RuO$_4$}
\newcommand{\caruoea}{Ca$_{1.8}$Sr$_{0.2}$RuO$_4$}
\newcommand{\caruoef}{Ca$_{1.5}$Sr$_{0.5}$RuO$_4$}

\newcommand{\casrruo}{Ca$_{2-x}$Sr$_{x}$RuO$_4$}

\newcommand{\gbd}{$\gamma$-band}

\newlength{\figwidth}

\divide\figwidth by 2

\begin{document}

\advance\vsize by 2 cm

\title{Structural aspects of metamagnetism in Ca$_{2-x}$Sr$_{x}$RuO$_4$ ($0.2 < x < 0.5$):
field tuning of orbital occupation}

\author{M.~ Kriener}
\affiliation{ II. Physikalisches Institut, Universit\"at zu
K\"oln, Z\"ulpicher Str. 77, D-50937 K\"oln, Germany}

\author{P. Steffens}
\affiliation{ II. Physikalisches Institut, Universit\"at zu
K\"oln, Z\"ulpicher Str. 77, D-50937 K\"oln, Germany}

\author{J. Baier}
\affiliation{ II. Physikalisches Institut, Universit\"at zu
K\"oln, Z\"ulpicher Str. 77, D-50937 K\"oln, Germany}

\author{O. Schumann}
\affiliation{ II. Physikalisches Institut, Universit\"at zu
K\"oln, Z\"ulpicher Str. 77, D-50937 K\"oln, Germany}

\author{T. Zabel}
\affiliation{ II. Physikalisches Institut, Universit\"at zu
K\"oln, Z\"ulpicher Str. 77, D-50937 K\"oln, Germany}

\author{T. Lorenz}
\affiliation{ II. Physikalisches Institut, Universit\"at zu
K\"oln, Z\"ulpicher Str. 77, D-50937 K\"oln, Germany}

\author{O. Friedt}
\affiliation{ II. Physikalisches Institut, Universit\"at zu
K\"oln, Z\"ulpicher Str. 77, D-50937 K\"oln, Germany}

\author{R. M\"uller}
\affiliation{ II. Physikalisches Institut, Universit\"at zu
K\"oln, Z\"ulpicher Str. 77, D-50937 K\"oln, Germany}

\author{A. Gukasov}
\affiliation{ Laboratoire L\'eon Brillouin, C.E.A./C.N.R.S.,
F-91191 Gif-sur-Yvette CEDEX, France}

\author{P. Radaelli}
\affiliation{ ISIS Facility, Rutherford Appleton Laboratory,
Chilton, Didcot, Oxon OX11 0QX, United Kingdom}
\affiliation{Department of Physics and Astronomy, University
College London, London WC1E 6BT, United Kingdom}

\author{P. Reutler}
\affiliation{II. Physikalisches Insitut, RWTH Aachen, Huykensweg,
D-52056 Aachen, Germany} \affiliation{Laboratoire de
Physico-Chimie de l'Etat Solide, Universit\'e Paris Sud, 91405
Orsay Cedex, France}

\author{A. Revcolevschi}
\affiliation{Laboratoire de Physico-Chimie de l'Etat Solide,
Universit\'e Paris Sud, 91405 Orsay Cedex, France}

\author{S.~Nakatsuji}
\affiliation{Department of Physics, Kyoto University, Kyoto
606-8502, Japan}

\author{Y.~Maeno}
\affiliation{Department of Physics, Kyoto University, Kyoto
606-8502, Japan} \affiliation{International Innovation Center and
Departement of Physics, Kyoto 606-8502, Japan\\}

\author{M. Braden}
\email{braden@ph2.uni-koeln.de}%
\affiliation{ II. Physikalisches Institut, Universit\"at zu
K\"oln, Z\"ulpicher Str. 77, D-50937 K\"oln, Germany}

\date{\today, \textbf{preprint}}


\begin{abstract}

The crystal structure of Ca$_{2-x}$Sr$_x$RuO$_4$ with $0.2<x<1.0$
has been studied by diffraction techniques and by high resolution
capacitance dilatometry as a function of temperature and magnetic
field. Upon cooling in zero magnetic field below about 25\,K the
structure shrinks along the $c$-direction and elongates in the
$a, b$ planes ($0.2<x<1.0$), whereas the opposite occurs upon
cooling at high-field ($x=0.2$ and 0.5). These findings indicate
an orbital rearrangement driven by temperature and magnetic
field, which accompanies  the metamagnetic transition in these
compounds.

\end{abstract}

\maketitle

The phase diagram of \csruo ~possesses quite different end
members with the spin-triplet superconductor \srruo ~on the one
side \cite{maeno} and the antiferromagnetically ordered Mott
insulator \caruo ~on the other side \cite{caruo1,caruo2}. Since
Sr and Ca are both divalent, one has to attribute the different
physical behavior \cite{3a,3b} entirely to the difference in the
ionic radii.
For small Ca content the octahedra present a $c$ axis rotation
and for higher Ca content ($x<0.5$) a tilt of the octahedra
around an in-plane axis occurs \cite{friedt}. These distortions,
through reduction of the hybridization, imply smaller band
widths, which together with a constant Hubbard type interaction,
enhance the correlation effects \cite{fang}. For Sr content lower
than 0.2, finally Mott localization occurs in a material
\cite{3a,3b} which exhibits a strong rotation and a strong tilt
deformation \cite{friedt}.

Outstanding physical properties are found in compounds in the
metallic regime but close to localization, 
$0.2<x<0.5$. At $T\sim 2$\,K samples with $x\sim 0.5$ exhibit a
magnetic susceptibility a factor of 200 higher than that of pure
\srruo ~\cite{satoru-prl}. In addition, the linear coefficient in
the specific heat is exceptionally high, of the order of
${C/T}\sim 250\,\frac{\rm mJ}{\rm mole\,K^2}$
\cite{satoru-prl,jin}, well in the range of typical heavy fermion
compounds. Inelastic neutron scattering has revealed strongly
enhanced magnetic fluctuations of incommensurate character
\cite{friedt-prl}, very different from those in pure \srruo\
\cite{sidis,braden2002}.

For Sr concentrations lower than 0.5, but still in the metallic
phase, the tilt distortion occurs and strongly modifies the
physical properties. The magnetic susceptibility at 2\,K, measured
in a low field, decreases with decreasing Sr content and
increasing tilt; for $0.2<x<0.5$ there is a maximum in the
temperature dependence of the susceptibility \cite{3b}. In this
concentration range the low-temperature low-field magnetization
is, hence, small compared with the extrapolation both from high
temperature and from high Sr concentration. A metamagnetic
transition occurs in these compounds at low temperature yielding
a high-field magnetization for $x=0.2$ which actually exceeds
that for $x=0.5$ \cite{satoru-prl}. A similar meta\-magnetic
transition has been reported for \srdzs ~\cite{sr327}, where the
related quantum critical end point has been proposed to cause
outstanding transport properties.


Single crystals of \casrruo ~were grown by a floating zone
technique in image furnaces at Kyoto University ($x=0.2$, 0.62 and
1.0) and at Universit\'e Paris Sud ($x=0.5$); a powder sample of
\caruoea ~was prepared by the standard solid state reaction at
Universit\"at zu K\"oln. Single crystal neutron diffraction
experiments were performed on the lifting counter diffractometer
6T.2 at the Orph\'ee reactor in magnetic fields up to 7\,T. Using
the GEM time of flight diffractometer at the ISIS facility,
powder diffraction patterns were recorded in fields up to 10\,T.
Thermal expansion and magnetostriction were studied in magnetic
fields up to 14\,T \cite{dilatometry}.

The thermal expansion was determined on the single crystals of
compositions, $x=0.2$, 0.5, 0.62, and 1.0, along the $c$ axis and
along an in-plane direction. The results are presented in Fig.\ 1
together with the relative length changes obtained by
integration. In all samples we find qualitatively identical
anomalies occurring near $T\sim 25$\,K; these are strongest in
\caruoea . There is a shrinking of the $c$ axis and an elongation
of the RuO$_2$ plane at low temperature. In contrast to these
low-temperature anomalies, the thermal expansion at higher
temperature is qualitatively different within the series. Samples
with $x>0.5$ show a normal positive thermal expansion along the
planes but a negative thermal expansion coefficient along $c$,
whereas the samples with smaller Sr content exhibit nearly the
opposite. 


\begin{figure}[tp]
\vskip 0.001 true cm
 \resizebox{0.65\figwidth}{!}{
\includegraphics*[angle=-0]{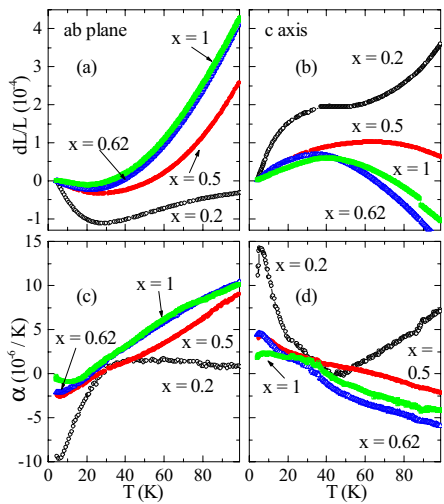}}
\vskip -0.1 true cm
 \caption{Temperature dependence of the integrated relative length changes a), b) and the thermal expansion coefficients
 c), d) for different Sr concentrations:
 a), c) and b), d) refer to the in-plane and to the $c$ directions, respectively.
 }\label{fig1}
\end{figure}

Whereas all high-temperature effects may be explained by the
structural arguments \cite{note-struc}, the thermal expansion
anomalies at low temperature must have an electronic origin. The
expansion coefficients exhibit extrema below 10\,K excluding an
explanation by anomalous phonon Gr\"uneisen parameters, since
there are no optical or zone-boundary modes in the energy range
of 1\,meV and below. Instead, the electronic Gr\"uneisen
parameter at low temperature appears to be extraordinarily large.

The thermal-expansion anomalies have been confirmed by diffraction
techniques, which may analyze both orthorhombic in-plane
directions independently, see Fig.\ 2. The low-temperature
expansion upon cooling is observed along both in-plane
directions, whereas the $a$ and $b$ directions show a different
thermal expansion at higher temperature due to the tilt-induced
orthorhombic distortion. The low-temperature anomalies are thus
not directly related to the tilt distortion. Nevertheless, the
fact that all features are strongest in the sample with strongest
tilt distortion suggests a pronounced coupling. The observed
expansion of the lattice along the planes and the compression
perpendicular to them qualitatively resembles the effect observed
at the metal-insulator transition in \caruo\ \cite{friedt}.
However, in \caruo , the changes of the lattice constants are
about a factor 30 -- 50 larger, and the electric resistivity
strongly increases below the structural change, whereas \caruoea
~stays metallic to the lowest temperatures \cite{3b}.
Nevertheless, the structural anomalies in \caruoea ~are
associated with the same electronic effect: a down shift of the
$d_{xy}$ orbital energy compared to those of the $d_{xz}$ and
$d_{yz}$ orbitals and, hence, a transfer of electrons from the
one-dimensional bands into the planar \gbd\ \cite{hao}(for
notation see \cite{mazin}).

For \caruoea , there is clear evidence that the anomalous
structural behavior at low temperature is accompanied by anomalies
in the magnetic and electronic properties detected earlier. In
Fig.\ 3a) -- c) the change in the lattice constants is compared to
the temperature dependencies of the low-field magnetization and of
the in-plane electric resistivity. The latter shows a down-turn
just around the temperature range where the thermal expansion
anomaly is observed. The magnetic susceptibility deviates from
Curie-Weiss behaviour in this temperature range, presenting a
maximum at about 10\,K \cite{3a,3b}. We conclude that the
structural anomaly in \caruoea ~is associated with an electronic
mechanism. In \caruoef ~there is a similar though much weaker
effect in the in-plane resistivity \cite{3a}; a related reduction
of the low-temperature magnetic susceptibility might be hidden by
its normal low-temperature increase.

 \begin{figure}[tp]
\resizebox{0.8\figwidth}{!}{
\includegraphics*{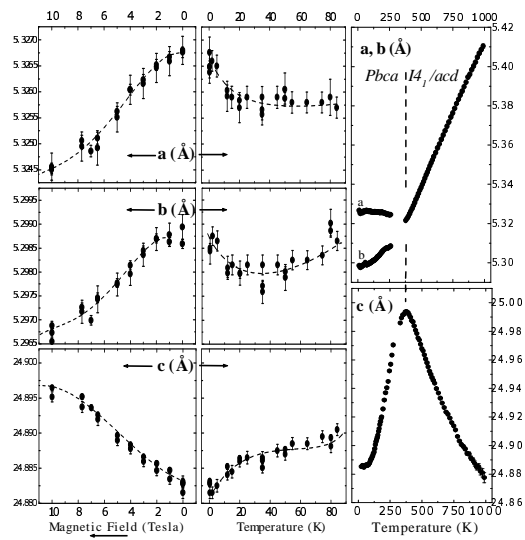}}
\vskip -0.01 true cm
 \caption{ Lattice parameters as a function of temperature and magnetic field:
 Right and middle: temperature dependence of the orthorhombic lattice parameters in \caruoea
 ~determined with a Laboratory X-ray diffractometer and the
 GEM diffractometer at ISIS. Left: magnetic field dependence of the lattice parameters
 obtained by neutron diffraction on GEM at 0.35\,K. }\label{fig2}
\end{figure}

The response to the metamagnetic transition in \caruoea ~was
studied by neutron diffraction, see Fig.\ 2,  by the measurement
of the magnetostriction with a high-resolution capacitance
dilatometer, Fig.\ 3d) and 3g), and by heat capacity studies ,
Fig.\ 3h) \cite{zabel}. In the magnetostriction experiment, see
Fig.\ 3d) and 3g), the field was oriented along the $c$ direction
and the length change was recorded parallel to the field. The
meta\-magnetic transition is clearly seen at 7\,T leading to an
enhancement of the $c$ axis by $\varepsilon _c =\frac{\Delta
L(B)}{L}\sim 6\times 10^{-4}$. In close resemblance with their
temperature dependencies, the structural change at the
metamagnetic transition is accompanied by changes in
magnetization and electrical resistivity. As shown in Fig.\ 3d)
-- f) magnetization, length change and $c$ axis magnetoresistivity
scale with each other. The field derivatives corresponding to the
magnetostriction and the field dependent susceptibility show a
similar field dependence as the in-plane magnetoresistivity
\cite{bemerkung2}. Also the specific heat over $T$ ratio,
$\frac{C_p}{T}$ \cite{zabel}, increases around the critical field
but much less. Comparing the susceptibility with $\frac{C_p}{T}$
quantitatively, one finds a normal Wilson ratio at low and high
fields, whereas the Wilson ratio at the metamagnetic transition
is strongly enhanced, close to 40, due to the much stronger
effect of the susceptibility. The maximum of the specific heat
coefficient at the metamagnetic field indicates that the system
transits between two different instabilities \cite{friedt-prl}. It
appears most likely that the metamagnetic transition in compounds
close to $x=0.5$ is shifted towards zero field; therefore, the
large susceptibility and the high Wilson ratio of \caruoef
~\cite{satoru-prl} at zero field may still be caused by the same
physics.

\begin{figure}[tp]
\vskip -0.5 true cm \resizebox{0.8\figwidth}{!}{
\includegraphics*{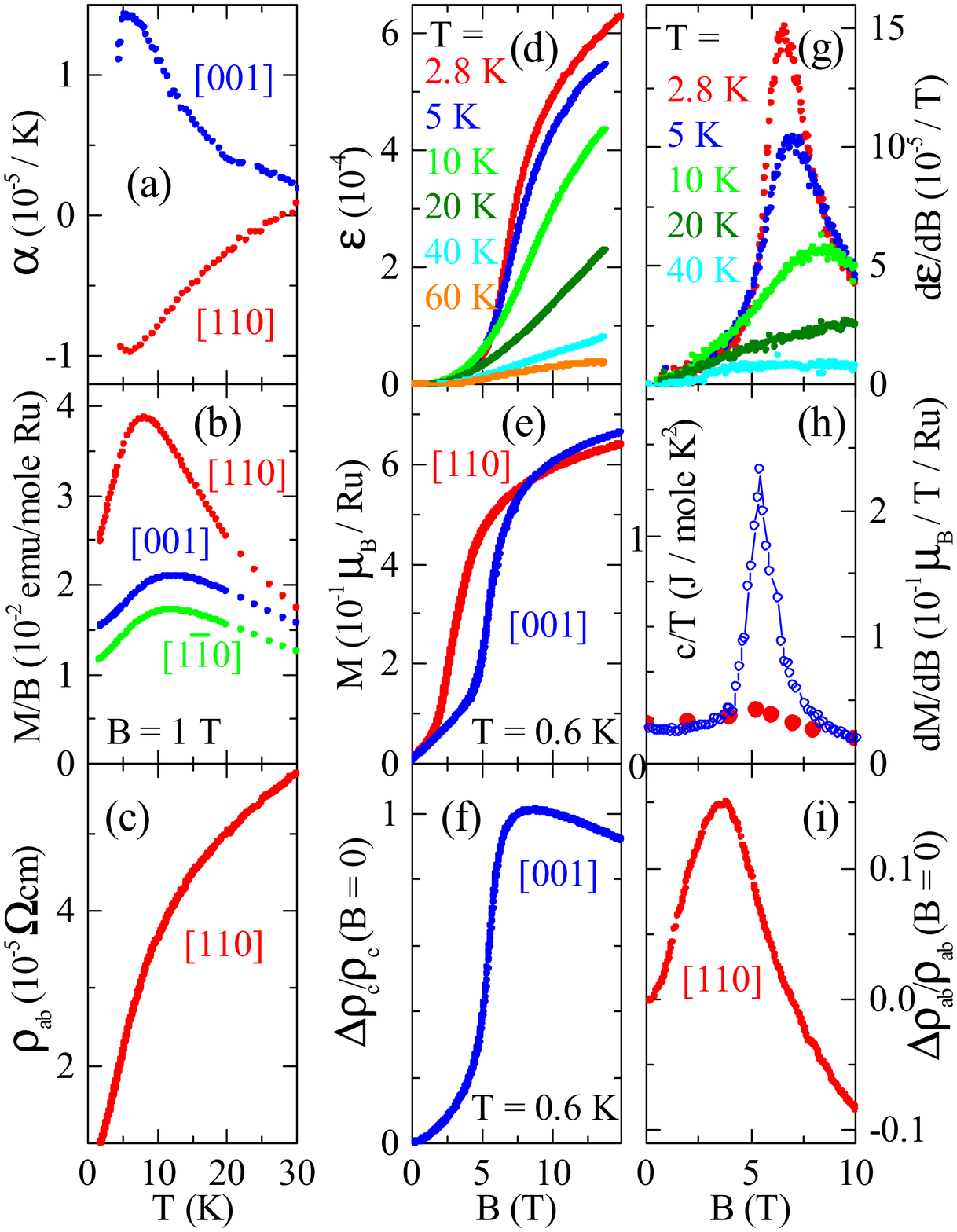}
} \vskip -0.1 true cm \caption{Comparison of the temperature
dependencies of a) thermal expansion, b) magnetization, and c)
in-plane resistivity for \caruoea. d) -- i) Comparison of the
magnetic-field dependencies in \caruoea: d) the longitudinal
magnetostriction $\varepsilon_c={\Delta}L/L)$ for a field along
the $c$ direction, e) the magnetization for fields along [110] and
[001] directions (note that the metamagnetic transition depends
on the orientation of the field), f) the magneto-resistivity
along $c$ direction, g) the field derivative of the
magnetostriction, h) the field derivative of the magnetization
and the linear coefficient of the specific heat $\frac{C_p}{T}$
\cite{zabel}, and i) the in-plane longitudinal
magneto-resistivity. Resistivity, magneto-resistivity and
magnetization data were taken from references
\cite{satoru-prl,3a,3b}.}\label{fig3}
\end{figure}

In the diffraction experiment the metamagnetic transition is
smeared out due to the random orientation of the grains with
respect to the field \cite{satoru-prl}. However, the overall
magnetostriction remains clearly visible, see Fig. 2. Crossing the
transition into the high-field phase we find a shrinking along
both in-plane directions and an enhancement along the $c$
direction. The absolute value of the magnetostriction along $c$
agrees well with the dilatometer result. These observations
indicate that the structural distortion occurring upon cooling in
zero-field may be suppressed and even be inverted by applying a
high field at low temperature. From the neutron powder
diffraction data obtained on GEM at ISIS we may deduce that the
crystal structure remains essentially unchanged with the field.
In particular, there is no evidence for superstructure
reflections which would appear or disappear with magnetic field.
By measuring the tilt superstructure reflection on a single
crystal with the lifting counter diffractometer 6T.2 at the
Orph\'ee reactor, we can precisely determine the tilt angle
reduction to only 3\,\% ~(under a field of 7\,T) in agreement
with the powder diffraction study on GEM (ISIS). Since the tilt
distortion is coupled with a shrinking of the $c$ axis and with
an increase of the averaged in-plane parameters, the observed
tilt reduction agrees well with the effect in the lattice
constants.

The comparison of the thermal expansion coefficients measured
with and without magnetic field is shown in Fig.\ 4 for \caruoea
~and \caruoef. The pronounced shrinking along the $c$ direction
in zero field is successively suppressed by the field and turns
into a low-temperature elongation at fields larger than 6\,T. In
contrast, the effect observed along the RuO$_2$ planes in \caruoef
~changes from a low-temperature elongation into a low-temperature
compression. All our observations indicate that there is an
electronic rearrangement occurring at low temperature which may be
tuned by the magnetic field. At zero field, electrons shift from
$d_{xz}$ and $d_{yz}$ orbitals into the $d_{xy}$ orbitals upon
cooling, whereas they do the opposite at high field. This thermal
effect is largely enhanced by the octahedron tilt.
%
The structural anomalies indicate a competition between different
electronic ground states similar to typical heavy fermion
compounds \cite{stewart} or LiV$_2$O$_4$ \cite{kondo,chmaissem}.
The low-magnetization state is most likely related to the magnetic
instability evidenced in the strongly enhanced magnetic
fluctuations in \caruoef\ \cite{friedt-prl}.

\begin{figure}
\vskip -0 true cm \resizebox{0.8\figwidth}{!}{
\includegraphics*{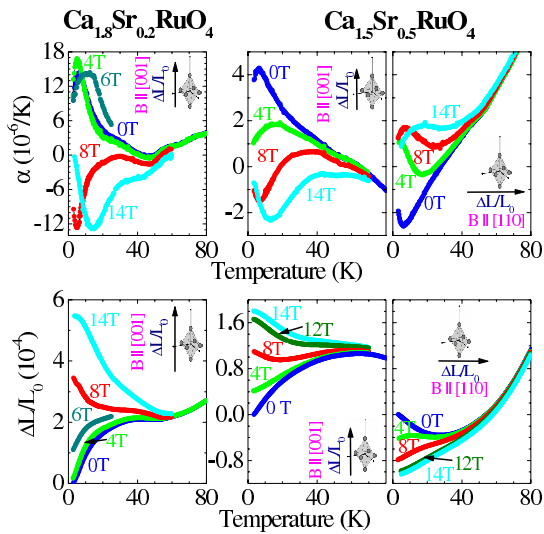}}
\vskip -0.02 true cm \caption{a) (Top) Thermal expansion and
integrated length changes (Bottom) determined at different
magnetic fields. The left and right parts present data for
\caruoea ~and \caruoef, respectively. In a) -- d) data were taken
parallel to the field applied along the $c$ direction, in e) the
field was applied parallel to the RuO$_2$ planes.} \label{fig4}
\end{figure}

The observed anomalies in the temperature dependence of the
lattice parameters clearly indicate a change in the orbital
occupation which may be tuned by the external field. We propose an
interpretation based on the van Hove singularity, vHs,  in the
\gbd  ~formed by the $d_{xy}$ orbitals \cite{mazin}. The vHs is
unoccupied in pure \srruo, but the electronic structure is
significantly changed by the rotational distortion which reduces
the width of the \gbd. LDA calculations and recent ARPES
measurements indicate that the vHs in \caruoef \ is occupied and
that the \gbd \ is hole-like \cite{fang,ding}, as it was also
deduced from the inelastic magnetic scattering \cite{friedt-prl}.
The zero-field thermal expansion anomalies are associated with a
transfer of electrons from the one-dimensional $d_{xz}$ and
$d_{yz}$ bands into the \gbd. This corresponds to a shift of the
DOS peak related to the vHs away from the Fermi level, as the vHs
singularity is below the Fermi-level in \caruoef \cite{ding}.
Since the high DOS is intrinsically related to ferromagnetism, the
shift may explain the reduction of magnetic susceptibility as a
function of both temperature and Sr content. In the tilted phase
the structural anomalies are strengthened due to some direct
coupling between the tilt and orbital occupation. The tilting
reduces the band width in the one-dimensional bands \cite{fang},
thereby increasing the local character of these particles, and
induces a further deformation of the RuO$_6$ octahedra, i.\,e.\ an
elongation perpendicular to the tilt axis \cite{friedt,3a}. Upon
increase of the magnetic field, the high DOS arrangement is
stabilized, thereby the structural distortions are reversed and
electrons are shifted from the \gbd ~to the one-dimensional
bands. The almost localized character of the electrons in this
concentration range rather close to the metal-insulator transition
must be very important, since the relatively small structural
changes are able to cause qualitatively different physical
behavior. Also, the occurrence of the zero-field thermal
expansion anomalies at low temperature may be understood in such
a picture only if the Fermi surface and Fermi liquid effects are
restricted to low temperature. In an usual metal the interplay
between the crystal structure and the electron system should be
determined at much higher temperature due to the high electronic
energy scale involved.

In conclusion, we find strong anomalies in the temperature and
field dependence of the crystal structure in \casrruo\
($0.2<x<0.5$) which indicate that the $t_{2g}$ orbital occupation
is tuned by Sr concentration, temperature and external magnetic
field.

{\bf Acknowledgments}  This work was supported by the Deutsche
Forschungsgemeinschaft through the Sonderforschungsbereich 608
and by Grants-in-Aid for scientific research from MEXT of Japan.
We are grateful to L. Balicas and D.\,Khomskii for interesting
discussions.


\begin{thebibliography}{28}
\expandafter\ifx\csname
natexlab\endcsname\relax\def\natexlab#1{#1}\fi
\expandafter\ifx\csname bibnamefont\endcsname\relax
  \def\bibnamefont#1{#1}\fi
\expandafter\ifx\csname bibfnamefont\endcsname\relax
  \def\bibfnamefont#1{#1}\fi
\expandafter\ifx\csname citenamefont\endcsname\relax
  \def\citenamefont#1{#1}\fi
\expandafter\ifx\csname url\endcsname\relax
  \def\url#1{\texttt{#1}}\fi
\expandafter\ifx\csname
urlprefix\endcsname\relax\def\urlprefix{URL }\fi
\providecommand{\bibinfo}[2]{#2}
\providecommand{\eprint}[2][]{\url{#2}}

\bibitem{maeno}\bibinfo{author}{\bibfnamefont{Y.}~\bibnamefont{Maeno et al.}},
  \bibinfo{author}{\bibfnamefont{Y.}~\bibnamefont{Maeno}},
  \bibinfo{journal}{J. of Sol. State Chem.}
  \textbf{\bibinfo{volume}{156}}, \bibinfo{pages}{26} (\bibinfo{year}{2001}).

\bibitem{caruo1}S. Nakatsuji et al., J. Phys. Soc. Jpn. 66, 1868
(1997).


\bibitem{caruo2}\bibinfo{author}{\bibfnamefont{M.}~\bibnamefont{Braden et al.}},
  \bibinfo{journal}{Phys.~Rev.~B} \textbf{\bibinfo{volume}{58}},
  \bibinfo{pages}{847} (\bibinfo{year}{1998}{\natexlab{a}}).



\bibitem{3a}\bibinfo{author}{\bibfnamefont{S.}~\bibnamefont{Nakatsuji et al.,}}
  \bibinfo{journal}{Phys.~Rev.~B} \textbf{\bibinfo{volume}{62}},
  \bibinfo{pages}{6458} (\bibinfo{year}{2000}{\natexlab{a}}).


\bibitem{3b}\bibinfo{author}{\bibfnamefont{S.}~\bibnamefont{Nakatsuji et al.,}}
  \bibinfo{journal}{Phys.~Rev.~Lett.} \textbf{\bibinfo{volume}{84}},
  \bibinfo{pages}{2666} (\bibinfo{year}{2000}{\natexlab{b}}).



\bibitem{friedt}\bibinfo{author}{\bibfnamefont{O.}~\bibnamefont{Friedt et al.}},
  \bibinfo{journal}{Phys.~Rev.~B} \textbf{\bibinfo{volume}{63}},
  \bibinfo{pages}{174432} (\bibinfo{year}{2001}).



\bibitem{fang}\bibinfo{author}{\bibfnamefont{Z.}~\bibnamefont{Fang}} \bibnamefont{et al.}
  \bibinfo{journal}{Phys.~Rev.~B} \textbf{\bibinfo{volume}{64}},
  \bibinfo{pages}{020509} (\bibinfo{year}{2001}).


\bibitem{satoru-prl}\bibinfo{author}{\bibfnamefont{S.}~\bibnamefont{Nakatsuji et al.}},
  \bibinfo{journal}{Phys.~Rev.~Lett.} \textbf{\bibinfo{volume}{90}},
  \bibinfo{pages}{137202} (\bibinfo{year}{2003}).

\bibitem{jin} R. Jin et al., cond-mat0112405.


\bibitem{friedt-prl} O. Friedt et al., cond-mat0311652


\bibitem{sidis}\bibinfo{author}{\bibfnamefont{Y.}~\bibnamefont{Sidis et al.}},
  \bibinfo{journal}{Phys.~Rev.~Lett.} \textbf{\bibinfo{volume}{83}},
  \bibinfo{pages}{3320} (\bibinfo{year}{1999}).

\bibitem{braden2002}
\bibinfo{author}{\bibfnamefont{M.}~\bibnamefont{Braden et al.}},
  \bibinfo{journal}{Phys.~Rev.~B} \textbf{\bibinfo{volume}{66}},
  \bibinfo{pages}{064522} (\bibinfo{year}{2002}{\natexlab{a}}).


\bibitem{sr327}S.A. Grigera et al., Science 294, 329 (2001); R.S.
Perry et al., Phys. Rev. Lett. 86, 2661 (2001).

\bibitem{dilatometry} R. Pott and R. Schefzyk, J. Phys. E16, 444
(1983); T. Lorenz et al., Phys. Rev. B 55, 5914 (1997).

\bibitem{note-struc} The increase of  the static tilt deformation causes a
shrinking of the $c$ parameter, see reference \cite{9}. Since
uniaxial pressure along $c$ stabilizes the tilt, it will cause a
frequency softening and a negative Gr\"uneisen parameter in the
non-tilted phase.


\bibitem{9} M. Braden et al.,
Z. Physik B 94, 29 (1994).


\bibitem{hao}T. Mizokawa et al.,
Phys. Rev. Lett. 87, 077202 (2001).


\bibitem{mazin}\bibinfo{author}{\bibfnamefont{D.~J.} \bibnamefont{Singh}},
  \bibinfo{journal}{Phys.~Rev.~B} \textbf{\bibinfo{volume}{52}},
  \bibinfo{pages}{1358} (\bibinfo{year}{1995});
\bibinfo{author}{\bibfnamefont{A.}~\bibnamefont{Liebsch}} \bibnamefont{and}
  \bibinfo{author}{\bibfnamefont{A.}~\bibnamefont{Lichtenstein}},
  \bibinfo{journal}{Phys.~Rev.~Lett.} \textbf{\bibinfo{volume}{84}},
  \bibinfo{pages}{1591} (\bibinfo{year}{2000});
\bibinfo{author}{\bibfnamefont{T.}~\bibnamefont{Oguchi}},
  \bibinfo{journal}{Phys.~Rev.~B} \textbf{\bibinfo{volume}{51}},
  \bibinfo{pages}{1385} (\bibinfo{year}{1995}).






\bibitem{stewart}S.H. Liu, Handbook on the Physics and Chemistry of Rare
Earths, Vol. 17, Elsevier Science Publishers, 87 (1993), G.R.
Stewart, Rev. Mod. Phys. 56, 755 (1984).

\bibitem{kondo}
\bibinfo{author}{\bibfnamefont{S.}~\bibnamefont{Kondo et~al.}},
  \bibinfo{journal}{Phys.~Rev.~Lett.}
  \textbf{\bibinfo{volume}{78}}, \bibinfo{pages}{3729} (\bibinfo{year}{1997}).


\bibitem{chmaissem}  \bibinfo{author}{\bibfnamefont{O.}~\bibnamefont{Chmaissem et al.}},
  \bibinfo{journal}{Phys.~Rev.~Lett.} \textbf{\bibinfo{volume}{79}},
  \bibinfo{pages}{4866} (\bibinfo{year}{1997}).


\bibitem{zabel} T. Zabel, PhD-thesis, Universit\"at zu K\"oln
(2004).

\bibitem{bemerkung2} The longitudinal $c$-axis magneto-resistivity
may be understood through the elongation of the $c$ axis, whereas
the maximum in-plane magnetoresistivity just at the field of the
metamagnetic transition suggests scattering with low-energy
fluctuations. Therefore, the $c$ and in-plane magneto-resistivity
scale with the magnetization and its field derivative,
respectively.

\bibitem{ding}S.-C. Wang et al., cond-mat/0407040.

\end{thebibliography}
\end{document}